\documentclass[conference]{IEEEtran}
\IEEEoverridecommandlockouts
\usepackage{cite}
\usepackage{amsmath,amssymb,amsfonts,amsthm}
\usepackage{algorithm}
\usepackage{graphicx}
\usepackage{multicol, blindtext}
\usepackage{textcomp}
\usepackage{svg}
\usepackage{subfig}
\usepackage{amsmath}

\DeclareMathOperator\erf{erf}
\graphicspath{ {figures/} }
\usepackage{color} 
\usepackage{float}
\usepackage{verbatim}

\usepackage[noend]{algpseudocode}

\def\BibTeX{{\rm B\kern-.05em{\sc i\kern-.025em b}\kern-.08em
T\kern-.1667em\lower.7ex\hbox{E}\kern-.125emX}}

\IEEEoverridecommandlockouts

\begin{document}

\title{\fontsize{23.7}{26}\selectfont
Standalone Deployment of a Dynamic Drone Cell for Wireless Connectivity of Two Services}

\author{
\IEEEauthorblockN{Igor Donevski, Jimmy Jessen Nielsen, Petar Popovski
\IEEEauthorblockA{Department of Electronic Systems, Aalborg University, Denmark}}
\IEEEauthorblockA{e-mails: igordonevski@es.aau.dk, jjn@es.aau.dk, petarp@es.aau.dk}

\thanks{This work has been submitted to IEEE for possible publication. Copyright may be transferred without notice, after which this version may no longer be accessible.}
} 
\maketitle

\begin{abstract}
We treat a setting in which two priority wireless service classes are offered in a given area by a drone small cell (DSC). Specifically, we consider broadband (BB) user with high priority and reliability requirements that coexists with random access machine-type-communications (MTC) devices. The drone serves both connectivity types with a combination of orthogonal slicing of the wireless resources and dynamic horizontal opportunistic positioning (D-HOP). We treat the D-HOP as a computational geometry function over stochastic BB user locations which requires careful adjustment in the deployment parameters to ensure MTC service at all times. Using an information theoretic approach, we optimize DSC deployment properties and radio resource allocation for the purpose of maximizing the average rate of BB users. While respecting the strict dual service requirements we analyze how system performance is affected by stochastic user positioning and density, topology, and reliability constraints combinations. The numerical results show that this approach outperforms static DSCs that fit the same coverage constraints, with outstanding performance in the urban setting. 
\end{abstract}

\begin{IEEEkeywords}
Unmanned Aerial Vehicles, Dynamic Drone Small Cell, 5G, eMBB, mMTC, Slicing
\end{IEEEkeywords}

\section{Introduction}
\label{introduction}
Due to their adeptness, drones or unmanned aerial vehicles (UAVs), have a growing importance in the world of communications as potential aid and substitution to classic cellular infrastructure \cite{mainsurvey,tutorial}. Aside extraordinary and unconventional applications, low or high altitude platforms (LAPs or HAPs) and drones as small cells (DSCs) are proven to provide good coverage through avoiding strong signal shadowing due to their high altitude to ground based user equipment \cite{MYPAPER, optlap, atg, otherantenna,3dplace}. Hence, DSC proliferation in the next generation of wireless service has the leverage to lower capital and operating expenditures (CAPEX \& OPEX) in developed countries by up to 52\% and 42\% respectively, through only diminishing site management costs and complexity \cite{mainsurvey}.
The coming generation of 3GPP and IEEE 802.11 communications calls for the coexistence of diverse and heterogeneous services within the same area that are served through the same interface in a concept known as radio access network (RAN) slicing. 5G defines three canonical service types with drastically different requirements, massive machine-type-communications (mMTC), enhanced mobile broadband (eMBB), and ultra reliable low latency communications (URLLC) \cite{ppslicing} \cite{otherslicing}, and for the upcoming IEEE 802.11be protocol, ultra high speed service links and reliable and low-latency communications \cite{wifi}. In this setting, drones can modify their spatial position in favor of improving channel conditions for priority users when serving two separate service categories.%
\begin{figure}[t]
\centering
\includegraphics[width=0.9\columnwidth]{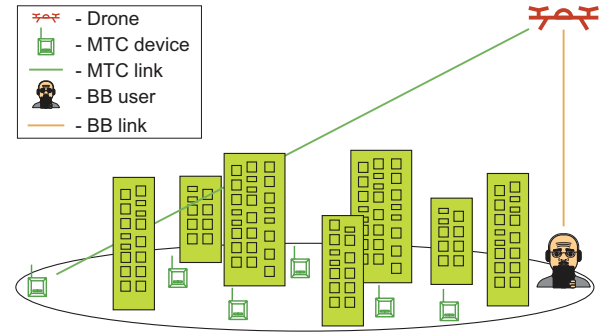}
\caption{Illustration of the drone provided MTC and BB links.}
\label{fig:toy}
\end{figure}
As illustrated in Fig. \ref{fig:toy} a dynamic standalone drone small cell (DSDSC) \cite{MYPAPER} exploits a dynamic horizontal opportunistic positioning (D-HOP) technique in strict priority for BB users, while successfully maintaining a lower tier reliability service MTC over the whole cellular area. In this setting, the drone's horizontal location can be treated as stochastic because the DSC relocates in favor of the BB users with stochastic positions and can assume extreme positions as the one illustrated.

\subsection{Relation to State of The Art}
The problem of adequate DSC positioning is recognized as cardinal in understanding deployment feasibility where the relative proportions of the DSC hold a key role in the performance of the system \cite{mainsurvey,tutorial,MYPAPER, optlap, atg, otherantenna,3dplace,babu2}. Additionally, exploiting the dynamic nature of DSCs to favor user locations is bound to improve system performance as per the works of \cite{MYPAPER, DCs}. The work in \cite{3dhetqos} extends the optimal drone positioning problem to serving multiple Quality of Service (QoS) categories. Regarding the provisioned traffic categories of 5G, the work of \cite{urllc} investigates the possibility of establishing the challenging traffic type of URLLC, to and from drones in different propagation environments. However, a common proposal towards slicing multiple services with DSCs is by employing tiered networks where drones are part of a complex architecture \cite{multier}. In accord, the work of \cite{hetopt} considers the UAV as modular equipment for more granular assistance to the needs of users with heterogeneous coverage demands. In a similar fashion, the works of \cite{cachingslice,UAVassisted} are concerned with keeping the slicing functionality of the cellular network with mixed special purpose UAVs. Finally, the emerging work of \cite{tquek} is concerned with maximizing the data rate for eMBB users while optimizing UAVs' total transmit power for drone provided slices. 

This work is concerned with providing reliable wireless coverage for areas that demand slicing of two traffic groups, that is enhanced by performing preferential movements to the prioritized group. This is a novel setting that has received a lack of attention especially for a drone that is alone in serving both broadband and random access traffic. Once deployed, the DSDSC provided interface is sliced between both MTC devices and BB users. Since appropriate deployment parameters assure optimal operation, our goal is maximizing the average rate for priority BB users, by jointly optimizing the slicing allocation per priority class and the DSC interdependent deployment properties of: DSC height, cell size, and antenna beamwidth. 

Mainly motivated by the information theoretic study of the case of slicing eMBB and mMTC in \cite{ppslicing}, this work conducts Monte Carlo simulations to express the complex probability mixture into the potential gains of such an implementation. The proposed model is the first one that to our knowledge considers stochastic horizontal DSDSC positions and is described in detail in Section \ref{model}. Followed by a detailed elaboration in maintaining dual coverage in the case of stochastic mobility and problem description in Section \ref{problem}. The simulation analysis and the adequate discussions are contained in Section \ref{numerical}. The final conclusions are drawn in Section \ref{conclusion}.
\section{System Model}

\begin{figure}[t]
\centering
\includegraphics[width=1\columnwidth]{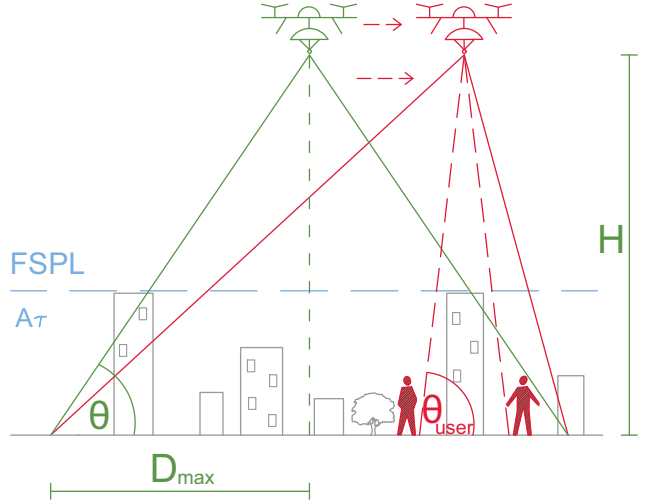}
\caption{ Vertical mapping of the propagation environment for reference drone position (green), and a snippet of DSDSC implementation with antenna tilting (red) \cite{MYPAPER}.}
\label{fig:model}
\end{figure}

To evaluate how DSC deployment impacts user service quality we assume an over-provisioned and stable back-haul. The cell is defined by a fixed circular area with radius $D_\text{max}$, that contains $K$ BB priority devices, where each $ i \in \{1,2,..,K\}$ is distributed by a Poisson Point Process (PPP) with intensity $\Lambda = A\lambda \text{ users}/\text{km}^\text{2}$. This makes $K$ a Poisson distributed random variable (RV) with density $\Lambda$ that has a squared relationship with $D_\text{max}$ as $A = \pi D_\text{max}^2$; and results in uniform BB user location distribution \cite{pfab}. 
As shown with red in Fig \ref{fig:model}, the drone position is defined by height $H$ with horizontal coordinates $(x_\text{d},y_\text{d})$, that can be off-center due to the operation of the D-HOP technique. The flying height of the drone $H$ is predefined during deployment and remains unchanged during D-HOP service since height modifications are incompatible with directional antenna use \cite{mainsurvey,tutorial,MYPAPER}.

Since our goal is to improve the BB service by best adjusting the wireless resources and drone deployment, we approach D-HOP from the perspective of solving a computational geometry problem over the random BB user locations. In this setting, the DSC has full knowledge of each priority device's ground location and activating D-HOP gives the DSC the ability to locate itself in the very center of the smallest bounding circle (SBC) that contains all the active BB users. The SBC is the most fairness oriented D-HOP approach \cite{MYPAPER} as it manages to achieve the shortest maximum horizontal distance from the drone to any node $D_\text{SBC}$. This renders BB user clusters to have no impact over the drone location. Since the number of BB users and their locations are stochastic, $D_\text{SBC}$ is a RV as well. 

In the rest of the paper, we express all horizontal distances between the drone and any point with horizontal coordinates $(x,y)$ as normalized by the cell radius $ \kappa  = \frac{ \sqrt{(x - x_\text{d})^2 + (y - y_\text{d})^2}}{D_\text{max} }$, where $0 \leq \kappa \leq 2$.
In the same way instead of $D_\text{SBC}$ we take its normalized version $\kappa_\text{SBC}= \frac{D_\text{SBC}}{D_\text{max}}$ when calculating BB service reliability.

\label{model}
\subsection{Propagation Model}
With green in Fig. \ref{fig:model} we illustrate the central position of the DSDSC that serves as a good reference through defining the reference elevation angle at the cell edge $\theta$. Shown with red in Fig. \ref{fig:model} we illustrate how a DSDSC may improve the propagation environment by avoiding large scale shadowing imposed by buildings that obstruct users' line of sight. To account for the positive impact of drone movements, we consider channel propagation impacted by large scale fading in which a user can belong to one of two propagation groups, users with line of sight (LoS) and no LoS (NLoS). In this way, the propagation losses come as a consequence of the free space path loss (FSPL) and excessive path loss \cite{optlap , atg}. To investigate the channel degradation for some user we need to know that user's elevation angle $\theta \leq \theta_{\text{user}} \leq \frac{\pi}{2}$ and formulate the path loss as:
%
\begin{multline}
\label{eq:Loss}
10\log(L) = - G_{t}(\theta) + 20\log(\theta D_\text{max}) \\ - 20\log(\sin(\theta_\text{user})) + 20\log(\frac{f4\pi}{c}) + A_{\tau}(\theta_{\text{user}} ),
\end{multline}
%
where: $\log$ is a shortened version of the common logarithm $\log_{10}$, $G_{t}(\theta)$ is the gain of the drone's antenna, the terms $20\log(\theta D_\text{max}) - 20\log(\sin(\theta_\text{user}))$ give the total direct air-to-ground distance between the drone and the user, the $\frac{f4\pi}{c}$ term is the FSPL term dependent on the operating frequency $f$, and finally, $A_{\tau}$ is a RV giving the excessive path loss value of large scale fading.

\subsubsection{Large Scale Fading} defined by $A_{\tau}$ is different for users belonging to either LoS and NLoS propagation group and we represent as:
\begin{equation}
\label{eq:bigA}
 A_{\tau}(\theta_{\text{user}}) = \tau A_{\tau = 1}(\theta_{\text{user}}) + (1 - \tau) A_{\tau = 0}(\theta_{\text{user}}),
\end{equation}
where $\tau = \{0,1\}$ is an indicator that can have value 1 with probability $ \text{P(LoS)} = \text{P}(\tau = 1)$ or 0 with probability $\text{P(NLoS)} = 1 - \text{P}(\tau = 1) = \text{P}(\tau = 0)$. The P(LoS) value is dependent on the elevation angle of the user $\theta_{\text{user}}$ given in radians, and topography dependent constants \textit{a} and \textit{b} \cite{atg}:
\begin{equation}
\label{eq:plos}
    \text{P(LoS)}  =\frac{1}{1+a\exp(-b(\theta_\text{user}\frac{180}{\pi}-a))}.
\end{equation}
The $A_{\tau}$ members are also probabilistic values that are Gaussian-Distributed with mean $\mu_\tau$ and variance dependent on the elevation angle of the user, and topology constants $d_\tau$ and $c_\tau$ for each propagation group $\tau$ as in \cite{atg}: 
\begin{equation}
\label{eq:manalph}
 A_{\tau}(\theta_{\text{user}}) \sim \textit{N}(\mu_{\tau},\sigma_{\tau}(\theta_{\text{user}})),
\end{equation}
where the variance is given as:
\begin{equation}
\label{eq:sigma}
 \sigma_{\tau}(\theta_{\text{user}}) = d_{\tau}\exp(-c_{\tau} \theta_{\text{user}}\frac{180}{\pi}).
\end{equation}

\subsubsection{The antenna Gain} $G_{t}(\theta)$ for the directional antenna mounted on the drone is taken as uniform gain over the whole aperture. The gain depends on the reference angle $\theta$, but is scaled by its effectiveness $ 0 < E_r < 1$ in matching an ideal antenna covering the same region, as given in \cite{MYPAPER}:
\begin{equation}
\label{eq:dt}
    G_{t}(\theta) = {E_r}10\log{(\frac{2}{1-\sin{(\theta)}})}.
\end{equation}
The drone tilts the antenna to cover the whole cellular area while moving, for which, the distortions of the circular shape to an oval one do not influence the analysis. 

\subsubsection{Path Loss Equivalent} $L_\text{e}(\theta,\kappa)$ is a reformulation of Eq. \eqref{eq:Loss} that represents the path loss experienced per meter of the cell's radius. In order to adapt this further to the case with stochastic users, we isolate the user positioning gain term and label it as $G_\text{U}(\theta_{\text{user}}) =A_{\tau}(\theta_{\text{user}}) - 20\log(\sin{(\theta_\text{user})})$, where $\theta_{\text{user}} = \arctan(\frac{\tan(\theta)}{\kappa})$. This leads to the final reformulation to get the path loss equivalence term:
\begin{multline}
\label{eq:lfinal}
L_\text{e}(\theta,\kappa) = 10\log(L) - 10\log(C) \\  = - G_t(\theta) + 20\log( \tan(\theta)) + G_\text{U}(\theta,\kappa),
\end{multline}
where: $C = \frac{f^24^2\pi^2{D_\text{max}}^2}{ c^2}$.

The path loss Equivalent metric is essential in establishing the coverage constraints and performance comparisons to classic DSC implementations that calculate the minimum path loss equivalent (MPLE). In detail, solving the MPLE problem gives the optimal $\theta$ for some fixed and specific normalized distance of $\kappa$; done as $\min_{\theta} L_\text{e}$. This approach ties all three deployment parameters $H$, $D_\text{max}$, and antenna beamwidth to be directly represented by the reference elevation angle $\theta$ and the radius of the cell $D_\text{max}$, where the latter has no impact on the MPLE \cite{MYPAPER, optlap , atg}. In contrast to this, $D_\text{max}$ does have an impact in choosing the optimal deployment parameters for the case of combined D-HOP and slicing which is addressed at the end of this Section. 
\subsection{Service Model}
We analyze the service requirements by considering that the rate provided to each user is given in bits per symbol. The number of users and scalability for each service can be addressed by bandwidth and power allocation methods that increase the number of symbols used for transmission on the channel, and are not relevant to the analysis. Therefore, the single interface is shared by a BB user and MTC user \cite{ppslicing} in an orthogonal multiple access (OMA) manner. Let $\omega_\text{B}$ where $0 \leq \omega_\text{B} \leq 1$ be the fraction of orthogonal time sharing for the BB user. Then the fraction left for the MTC user is $\omega_m = 1 -\omega_B$. Due to orthogonality, there is no interference between the services, such that the achievable rate $R_x$ for service BB and MTC where $x \in \{B,m\}$ respectively, is:
\begin{equation}
\label{eq:rate}
R_x = \omega_x \log_2(1+\frac{P_\text{t}  }{N C 10^{L_\text{e,x}/10}}),
\end{equation}
where $P_\text{t}$ is the transmit power, $N$ is the noise power, and both $C$ and $L_{e,x}$ represent signal attenuation as given in Eq. \eqref{eq:lfinal}.
\subsubsection{Service Reliability} for service $x$ can be calculated from probability of the rate of our system to be smaller than a required rate $R_x^\text{req}$. To find the $1 - \epsilon_x$ reliability \cite{ppslicing,urllc}, we first look at $P(R_x>R_x^\text{req})$. Using intermediary value for targeted user positioning gain $G_x$ as:
\begin{equation}
\label{eq:refgain}
G_x = - 10\log( (2^\frac{R_x^\text{req}}{\omega_x} - 1)\frac{N C}{ P_t} ) + G_{t}(\theta) - 20\log(\tan(\theta) ),
\end{equation}
we seek if it satisfies $G_\text{U}(\theta,\kappa) < G_x$ for the distribution of $A_{\tau}(\theta_{\text{user}})$ with system performance of $R_x>R_x^\text{req}$. We then calculate the Cumulative Density Function as:
\begin{align}
\label{eq:finalcdf}
\text{P}(G_\text{U}(\theta,\kappa) &< G_x) = \\
= \text{P}(\tau = 0)\text{P}(A_{\tau = 0} < u) &+ \text{P}(\tau = 1)\text{P}(A_{\tau = 1} < u),
\end{align}
where $u = G_x + 20\log(\sin{(\arctan(\frac{\tan(\theta)}{\kappa})})) $. The sum then unwraps into:
\begin{equation}
\small
\label{eq:expandedcdf}
\begin{split}
\text{P}(\tau = 0) &\int_{-\infty}^{u} \frac{1}{\sqrt{2\pi\sigma_{0}^2}} e^{- \frac{(x - \mu_{0} )^2}{2\sigma_{0}^2} } dx \\ 
+ \text{P}(\tau = 1) &\int_{-\infty}^{u} \frac{1}{\sqrt{2\pi\sigma_{1}^2}} e^{- \frac{(x - \mu_{1})^2 }{2\sigma_{1}^2} } dx = 
\\ 
\frac{1-\text{P}(\tau = 1)}{2}&(1+\erf( \frac{ u - \mu_0}{d_{0}\exp(-c_{0} \arctan(\frac{\tan(\theta)}{\kappa}\frac{180}{\pi})) \sqrt{2} } )
\\ 
+ \frac{\text{P}(\tau = 1)}{2}&(1+\erf( \frac{u - \mu_1}{d_{1}\exp(-c_{1} \arctan(\frac{\tan(\theta)}{\kappa}\frac{180}{\pi})) \sqrt{2} } ),
\end{split}
\end{equation}
where $\erf$ is the error function and the probability $\text{P}(\tau = 1) = \text{P(LoS)}$ is  taken as defined in Eq. \eqref{eq:plos}. An explicit expression for $\text{P}(G_\text{U}(\theta,\kappa)<G_x) = 1 -\epsilon_x$ in terms of $G_x$ is not possible and requires numerical solutions especially when calculating probability mixtures with $\kappa$ as a RV.
\subsubsection{Reference SNR of Static LAP} is needed to define a baseline testing scenario by mitigating the dependence on $D_\text{max}$. For this, we go back to the reference scenario where the drone is static in the center of the cell and solve MPLE for the cell edge $\kappa=1$, which is trivial and identical to the classic LAP \cite{optlap} case. The result is a pathloss equivalent named $L_\text{e-LAP}$ defined by placement at an optimal LAP cell edge elevation angle $\theta_\text{LAP}$. Since BB service is demanding $P( R_\text{B} \geq R_\text{B}^\text{req})=1-\epsilon_B$, we must be able to satisfy the path loss equivalent $L_\text{e-LAP}$ when we transmit with $P_\text{max}$ given by signal-to-noise ratio:
\begin{equation}
\label{eq:Pmax}
 \frac{P_\text{max} }{N} =  (2^{\frac{ R_\text{B}^\text{req}}{\omega_\text{B}}}-1)10^{L_\text{e-LAP}/10}C.
\end{equation}
This stands as the most basic application of a DSC \cite{optlap}, and it is expected that D-HOP enabled systems should outperform it when using the same power system driven by the identical SNR requirements of $\frac{P_\text{max}}{N}$. Finally, this action allows us to isolate the issue of the size of the cell with radius $D_\text{max}$ as follows in Section \ref{problem}.

\section{Problem analysis and Coverage Constraints}
\label{problem}

\begin{figure}[t]
\centering
\includegraphics[width=1\columnwidth]{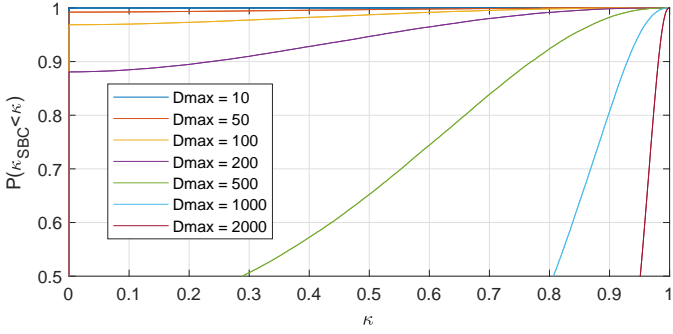}
\caption{Cumulative Probability Distribution Function of $\kappa_\text{SBC}$ for  $\lambda=2 \text{ users}/\text{km}^\text{2}$ in seven different cell sizes given in meters [m].}
\label{fig:cdf}
\end{figure}
To avoid misleading the reader that the drone offers perfect service for the priority users when no service is needed at all, we only consider the cases where $K>0$. An example of the $\kappa_\text{SBC}$ distribution in seven different cell sizes for $\lambda=2$ is shown on Fig \ref{fig:cdf}, from which we can conclude that increasing the cell radius affects the SBC algorithm to have less certainty in providing low $\kappa$ distances.
\subsection{BB Coverage}
In an active D-HOP implementation, the worst case distance for two or more BB users is given by $\kappa_\text{SBC}$ which is crucial in defining BB coverage. The rate at the edge of the SBC is given by $R_\text{SBC}(\theta,\kappa_\text{SBC})$, to which we define the BB coverage constraint as:
\begin{equation}
\label{eq:BBcov1}
\text{P}(\omega_B \log_2(1+\frac{P_\text{max}  }{N C 10^{L_\text{e,B}(\theta, \kappa_\text{SBC})/10}}) \geq R_\text{B}^\text{req}) \geq 1-\epsilon_B,
\end{equation}
where the power to noise ratio can be calculated from \eqref{eq:Pmax} and gives the constraint simplified as: 
\begin{equation}
\label{eq:BBcov}
\text{P}(  L_\text{e,B} \leq L_\text{e-LAP} )\geq 1-\epsilon_B.
\end{equation}
The complexity of calculating this constraint comes from the mixture distribution $G_\text{U}(\theta,\kappa_\text{SBC})$ in $L_\text{e,B}(\theta, \kappa_\text{SBC})$ that is given when passing $\kappa_\text{SBC}$ trough Eq. \eqref{eq:finalcdf}; an operation that must be done numerically.
\subsection{MTC Coverage}
The second tier MTC service should be offered to the whole cell, at all times, no matter where the drone is positioned. This is done since MTC access times are random and it is unrealistic to expect that we know the devices' positions with regards to the drone's position. We therefore account for the worst case MTC horizontal distance of $\kappa = 2$ when the drone is positioned at the edge of the cell. For a defined reference angle $\theta$, to guarantee the MTC service we first need to calculate the rate as in Eq. \eqref{eq:rate}. Since we set a power limit as in the reference LAP case we reach: 

\begin{equation}
\label{eq:pmtc}
\text{P}((1-\omega_B) \log_2(1+\frac{P_\text{max}  }{N C 10^{L_\text{e,m}(\theta, \kappa=2)/10}}) \geq R_\text{m}^\text{req}) \geq 1-\epsilon_m.
\end{equation}
Substituting the maximum power requirement for the reference LAP case from Eq. \eqref{eq:Pmax} we get the CDF for MTC coverage:
\begin{equation}
\label{eq:derivemtc}
 (2^{\frac{ R_\text{m}^\text{req}}{1-\omega_\text{B}}}-1)10^{L_\text{e}(\theta,\kappa=2)/10}C  \leq   (2^{\frac{ R_\text{B}^\text{req}}{\omega_\text{B}}}-1)10^{L_\text{e-LAP}/10}C ,
\end{equation}
which we can input in the final coverage constraint of:
\begin{equation}
\label{eq:a2}
\text{P}(L_\text{e-LAP} - 10\log(\frac{(2^{\frac{ R_\text{m}^\text{req}}{1-\omega_\text{B}}}-1)}{(2^{\frac{ R_\text{B}^\text{req}}{\omega_\text{B}}}-1)}) \geq  L_\text{e}(\theta,\kappa=2)) \geq 1-\epsilon_m.
\end{equation}

It should be noted that in this way, we are massively overprovisioning the MTC link when compared to the classic LAP case. This applies as we support MTC in the absolute worst case scenario of $\kappa$ which is necessary for guaranteeing random access.
\subsection{Maximal RAN slicing ratio}
If in the case of Eq. \eqref{eq:derivemtc} we calculate the MPLE value of $L_\text{e}(\theta,\kappa=2)$ we get some minimal value that can help us place a constraint for achieving maximum $\omega_\text{B}$. In other words, if we choose maximum possible value for $\omega_\text{B}$, there is only one $\theta$ at which the drone can operate and maintain MTC service for the whole cell. Therefore, using non-maximal $\omega_\text{B}$ should result in bigger range of possible $\theta$ values, at which MTC service is maintained. Finding the optimal $\theta$, can be done numerically through finding MPLE, and reach to a value of minimal path loss for $\kappa=2$ named, $L_\text{e-RSR}(\theta_\text{RSR},\kappa=2)$. And in order for MTC coverage to be satisfied, the following equation must hold:
%
%
\begin{equation}
\label{eq:maxw}
 10^{\frac{L_\text{e-LAP} -  L_\text{e-RSR}}{10}} \geq \frac{(2^{\frac{ R_\text{m}^\text{req}}{1-\omega_\text{B}}}-1)}{(2^{\frac{ R_\text{B}^\text{req}}{\omega_\text{B}}}-1)} .
\end{equation}
This ratio is an upper bound constraint for the value $\omega_\text{B}$, with which we can test if the scenario is within the theoretical limits of our system.
\subsection{Problem Description}
In a specific snapshot of the cell, the active BB user $i \in 1,2,3 \ldots K$, has a normalized horizontal distance $\kappa_i^\text{D}$ to the D-HOP activated drone, and a normalized horizontal distance of $\kappa_i^\text{S}$ to the scenario with a static LAP at the center of the cell. When served by the D-HOP enabled drone, this BB user will experience a rate of:
\begin{equation}
\label{eq:RD}
R_{i}^{\text{D}}(\theta,\omega_\text{B}) = \omega_\text{B} \log_2(1+ (2^{\frac{ R_\text{B}^\text{req}}{\omega_\text{B}}}-1)10^{L_\text{e-LAP}/10 - L_\text{e}(\theta,\kappa_i^\text{D})/10}) ,
\end{equation}
while in the reference LAP system located in the very center of the cell, the same BB user $i$ will experience a rate of:
\begin{equation}
\label{eq:RS}
R_{i}^{\text{S}}(\omega_\text{B}) = \omega_\text{B} \log_2(1+ (2^{\frac{ R_\text{B}^\text{req}}{\omega_\text{B}}}-1)10^{L_\text{e-LAP}/10 - L_\text{e}(\theta_\text{LAP},\kappa_i^\text{S})/10}) .
\end{equation}
Since our goal is to maximize the average rate for all active BB users in the cell we introduce the problem as:

\begin{IEEEeqnarray}{rCl}
\underset{\left\lbrace \theta, \omega_\text{B} \right\rbrace}{\text{maximize}} \,\,\,\, & &  \frac{E[R_{i}^{\text{D}}(\theta,\omega_\text{B})]}{E[R_{i}^{\text{S}}(\omega_\text{B})]}\label{eq:optcrit}\\
\text{s.t.} \,\,\,\, & & 0 < \theta < \frac{\pi}{2}\\
& &0 < \omega_\text{B}\\
& & \eqref{eq:BBcov}, \eqref{eq:a2}, \eqref{eq:maxw}
\end{IEEEeqnarray}

The objective function \eqref{eq:optcrit} is normalized by \eqref{eq:RS} because an analysis in the non-normalized case would not give productive results as the transmission power given by \eqref{eq:Pmax} scales with the reference deployment. Furthermore, the real altitude $H$ will directly depend on the deployment radius and not only $\theta$. This would then impact $D_\text{max}$, and in order to cover bigger areas, a multi drone provided coverage may be necessary such as in \cite{otherantenna,babu2}. Therefore, we are only interested in finding an optimal reference elevation angle $\theta$ and $\omega_\text{B}$ combination while we try to maximize the average rate improvements to the active BB users. Since the involved functions are monotonic within the given constraints, to efficiently find the maximum  we applied a binary-search algorithm to the Monte Carlo simulated estimates of $\kappa_\text{SBC}$ used to realize the $G_\text{U}$ mixture.
\section{Simulation and Results}
\label{numerical}
\begin{table}[t]
\caption{The excessive path loss parameters for: Suburban (1), Urban (2), and High Urban (3) environments \cite{atg}.}
\label{table:1}
\centering
\begin{tabular}{|l|l|l|l|l|l|l|l|l|}\hline
           & $a$    & $b$    & $\mu_{1}$   & $\mu_{0}$   & $d_{1}$ & $d_{0}$ & $c_{1}$ & $c_{0}$ \\\hline
1 & 4.88 & 0.43 & 0.1 & 21.0 &  11.25 & 32.17  &  0.06 & 0.03  \\\hline
2 &  9.61 & 0.16 &  1.0   &  20.0    &  10.39 & 29.6  & 0.05  & 0.03  \\ \hline
3 &  12.08 &  0.11 &  1.6   &   23.0   & 8.96  & 35.97  &  0.04  &   0.04 \\ \hline
\end{tabular}
\end{table}
We take a snapshot approach where at each snip the drone is at the SBC center of the BB users while maintaining the full MTC coverage as per the defined constraints. Each snip has no correlation to a previous one and no mobility of the users or trajectories are concerned. The service requirements for both priority groups are tested as per the eMBB and mMTC slicing requirements taken from \cite{ppslicing}. The mMTC devices require a rate of $R_\mathrm{m}^\text{req} = 0.3$ with low reliability of $\epsilon_m = 1-0.9$, and the rate for serving an eMBB user as $R_\mathrm{B}^\text{req} = 1$. In addition to medium $\epsilon_\mathrm{B} = 1-0.999$ (as in \cite{ppslicing}), we test under high $\epsilon_\mathrm{B} = 1-0.9999$ and ultra $\epsilon_\mathrm{B} = 1-0.99999$ eMBB reliability requirements.  We investigate this by testing the previously shown seven different cell sizes of $D_\mathrm{max} = [10\text{m},50\text{m},100\text{m},200\text{m},500\text{m},1000\text{m},2000\text{m}]$ in an area with fixed user density of $\lambda= 2 \text{ users}/\text{km}^\text{2}$. Additionally, we test three different deployment environments for a 2~GHz carrier, that affects $A_\tau$ as shown in Table \ref{table:1}. The topological constants required for calculating Eq. \eqref{eq:finalcdf} are taken from the Suburban, Urban and High Urban parameters of \cite{atg} from where High Urban is renamed from Dense Urban in \cite{atg} to avoid confusions regarding user density.
\begin{figure}[t]
\centering
\includegraphics[width=1\columnwidth]{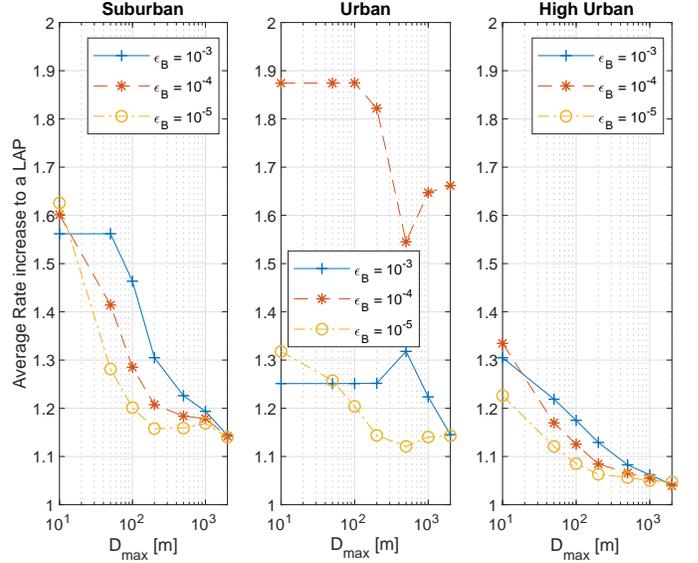}
\caption{ARI gain results for $E_\text{r} = 0.6$.}
\label{fig:rateperf}
\end{figure}

In Fig. \ref{fig:rateperf} we notice that the deployment feasibility, measured by the average rate increase (ARI), depends strongly on the local topology, where, Suburban and High Urban environments show that there is a strong decrease in D-HOP ARI in user-dense environments with an exception to this rule for the Urban deployments. In the case of High eMBB reliability for Urban deployments, the implementation achieves astonishing performance for all types of user densities, amounting to average rate improvements of 50-90\%. This performance is credited to the nature of the fading distribution, and the reliability combination for both eMBB and mMTC users occur very close to a point where the P(LoS) equation \eqref{eq:plos} shows a strong impact in the distribution of Eq. \eqref{eq:finalcdf}. Moreover, the DSDSC allows itself to serve mostly LoS links due to its mobility, while the static LAP has to meet the high reliability requirement of mostly, NLoS traffic. The average gains are subject to the much stricter reliability requirements of Eq. \eqref{eq:BBcov}, which results in optimal $\theta$ and $\omega_\text{B}$ combination limited by the distribution of the smallest SBC radius. Since $\kappa_\text{SBC}$ is directly dependent on the $D_\text{max}$ we see that ARI drops off steadily with the increase in cell size. Due to the fact that we measure performance as relative gain over a LAP equivalent setup and maintain cell proportions with $\theta$, the absolute dimensions of the cell do not have a direct impact on the measured rate but only impacts the user density distribution. Finally, by testing the four different $E_\text{r} = [0.1, 0.3, 0.6, 0.9]$ we only confirmed that better antenna compatibility results in overall higher optimal $\theta$, which in turn diminishes the advantages of D-HOP implementations and results in lower ARI \cite{MYPAPER}. When necessary to avoid clutter, we only show the performance for $E_\text{r} = 0.6$. 

\begin{figure}[t]
\centering
\includegraphics[width=1\columnwidth]{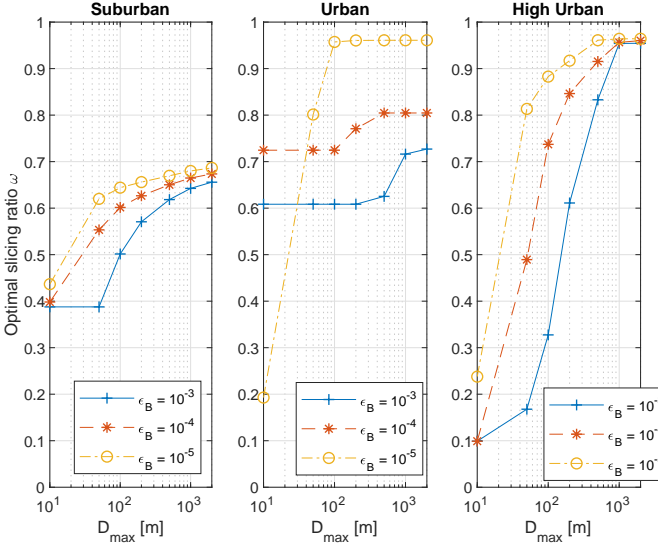}
\caption{Optimal Slicing Ratio $\omega_\text{B}$ evolution for $E_\text{r} = 0.6$}
\label{fig:wss}
\end{figure}
\begin{figure}[t]
\centering
\includegraphics[width=0.96\columnwidth]{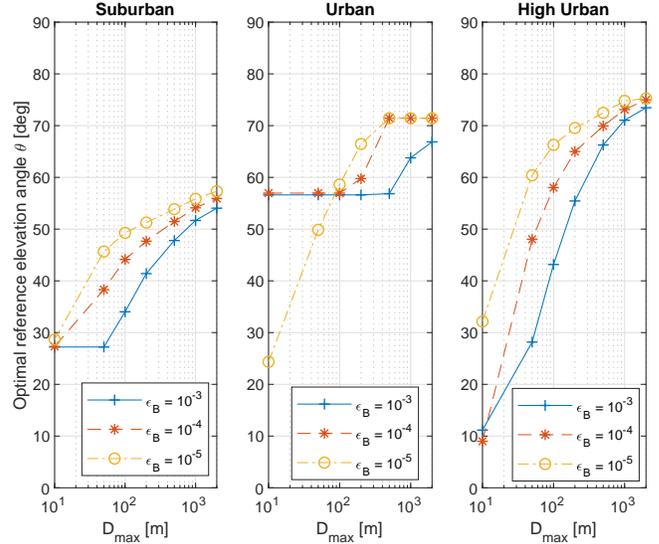}
\caption{Reference elevation angle $\theta$ evolution for $E_\text{r} = 0.6$}
\label{fig:angles}
\end{figure}

In our optimization problem the value $\omega_\text{B}$ fulfills two very important roles as it controls the rate and impacts the availability of $\theta$ for the optimization algorithm. The slicing ratios given in Fig. \ref{fig:wss} show strong correlation between low slicing ratio for eMBB users and high ARI performance. This is due to the relaxation of the mMTC coverage constraint for dynamic system coverage given in Eq. \eqref{eq:a2}. By relaxing this value, the SBC distribution allows for achieving lower operating optimal $\theta$ as shown in Fig. \ref{fig:angles} which places it closer to the eMBB user and results in higher ARI. 

\begin{figure}[b]
\centering
\includegraphics[width=1\columnwidth]{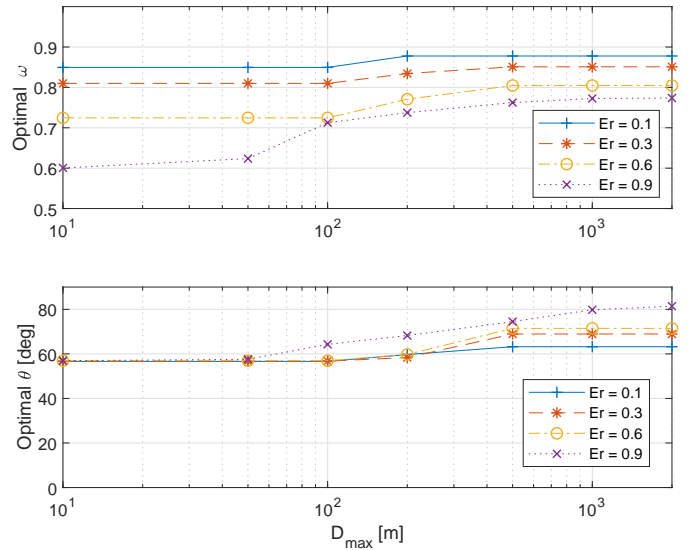}
\caption{Full $E_\text{r}$ analysis for $\epsilon_B = 1-0.9999$ in Urban deployment}
\label{fig:neww}
\end{figure}

We go back to analyze the most interesting case, the urban DSDSC implementation for high eMBB reliability as it shows strong performance advantages. The reason for this high outlier in the performance of this system can be deduced from Fig. \ref{fig:neww}. Here we can notice that the optimal $\theta$ for the system does not dramatically vary with neither user density nor antenna efficiency. Comparing the optimal slicing ratios, we notice that because of the point of where we are in the distribution, the system benefits very little from lowering the slicing ratio when compared to the other deployment cases. Hence, the optimal $\omega_\text{B}$ in this case is preferred to be lower for higher $E_\text{r}$ values, which is inverted from the other deployments. This means that the operating $\theta$ where the topology allows for the maximum slicing ratio constraint is very close to the optimal $\theta$ for the static system. In these cases, the slicing ratio is used to compensate for the differences in the use of more efficient antennas, and use the same optimal $\theta$. Such systems find themselves in a \textit{just right} or Goldilocks deployments scenario for the support of both schemes. Which in our case occurs in the urban scenario for slicing support of highly reliable eMBB traffic and low mMTC reliability support.
\section{Conclusion}
\label{conclusion}
This particular implementation of DSDSCs in slicing of MTC and BB traffic was tested in strict priority for BB users with coverage requirements for both services. Through Monte Carlo analysis we deduced that the use of this system outperforms its static counterpart in many different deployment scenarios, with expected superiority in implementations with low antenna efficiency and sparse BB user activation distributions. However, the implementation showed strong performance in the slicing DSDSC deployment in Urban environments for high reliability requirements for the BB traffic users. Opposed to the other deployment scenarios, this shows significant superiority with regards to static LAP small cells with average rate improvements of 50-90\% for the case of antenna with an efficiency coefficient of $E_r = 0.6$ . This analysis strongly highlights the importance of beforehand knowing the propagation properties of the environment as central in discovering the feasibility of DSDSC slicing deployments. We finally conclude that as per the given results, standalone deployment of a service slicing DSC can strongly benefit from dynamic and preferential horizontal movements. Finally, this research is limited in that it does not account for the times for which a drone would need to reach the new position, a topic reserved for a future iteration of the work.

\section*{Acknowledgment}
The work was supported by the European Union's research and innovation programme under the Marie Sklodowska-Curie grant agreement No. 812991 ''PAINLESS'' within the Horizon 2020 Program.


\begin{thebibliography}{10}

\bibitem{mainsurvey}
A. Fotouhi et al., "Survey on UAV Cellular Communications: Practical Aspects, Standardization Advancements, Regulation, and Security Challenges," in IEEE Communications Surveys \& Tutorials 2019.

\bibitem{tutorial}
M. Mozaffari, W. Saad, M. Bennis, Y. Nam and M. Debbah, "A Tutorial on UAVs for Wireless Networks: Applications, Challenges, and Open Problems," in IEEE Communications Surveys \& Tutorials, vol. 21, no. 3, pp. 2334-2360, thirdquarter 2019.

\bibitem{MYPAPER}
I. Donevski, J.J. Nielsen, "Dynamic Standalone Drone-mounted Small Cells" 2020 European Conference on Networks and Communications (EuCNC) , Dubrovnik, Croatia. IEEE, 2020.

\bibitem{optlap}
A. Al-Hourani, S. Kandeepan and S. Lardner, "Optimal LAP Altitude for Maximum Coverage," in IEEE Wireless Communications Letters, vol. 3, no. 6, pp. 569-572, Dec. 2014.

\bibitem{atg}
A. Al-Hourani, S. Kandeepan and A. Jamalipour, "Modeling air-to-ground path loss for low altitude platforms in urban environments," 2014 IEEE Global Communications Conference, Austin, TX, 2014, pp. 2898-2904.

\bibitem{3dplace}
M. Alzenad, A. El-Keyi, F. Lagum and H. Yanikomeroglu, "3-D Placement of an Unmanned Aerial Vehicle Base Station (UAV-BS) for Energy-Efficient Maximal Coverage," in IEEE Wireless Communications Letters, vol. 6, no. 4, pp. 434-437, Aug. 2017.

\bibitem{otherantenna}
M. Mozaffari, W. Saad, M. Bennis and M. Debbah, "Efficient Deployment of Multiple Unmanned Aerial Vehicles for Optimal Wireless Coverage," in IEEE Communications Letters, vol. 20, no. 8, pp. 1647-1650, Aug. 2016.

\bibitem{babu2}
N. Babu, C. B. Papadias and P. Popovski, "Energy-Efficient 3D Deployment of Aerial Access Points in a UAV Communication System," in IEEE Communications Letters,

\bibitem{ppslicing}
Popovski, Petar, et al. "5G wireless network slicing for eMBB, URLLC, and mMTC: A communication-theoretic view." IEEE Access 6 (2018): 55765-55779.

\bibitem{otherslicing}
X. Shen et al., "AI-Assisted Network-Slicing Based Next-Generation Wireless Networks," in IEEE Open Journal of Vehicular Technology, vol. 1, pp. 45-66, 2020.

\bibitem{wifi}
López-Pérez, David, et al. "IEEE 802.11 be Extremely High Throughput: The Next Generation of Wi-Fi Technology Beyond 802.11 ax." IEEE Communications Magazine 57.9 (2019): 113-119.

\bibitem{DCs}
B. Galkin, J. Kibiłda and L. A. DaSilva, "A Stochastic Model for UAV Networks Positioned Above Demand Hotspots in Urban Environments," in IEEE Transactions on Vehicular Technology, vol. 68, no. 7, pp. 6985-6996, July 2019.

\bibitem{3dhetqos}
Alzenad, Mohamed, Amr El-Keyi, and Halim Yanikomeroglu, "3-D placement of an unmanned aerial vehicle base station for maximum coverage of users with different QoS requirements." IEEE Wireless Communications Letters 7.1 (2017): 38-41.

\bibitem{urllc}
C. She, C. Liu, T. Q. S. Quek, C. Yang and Y. Li, "Ultra-Reliable and Low-Latency Communications in Unmanned Aerial Vehicle Communication Systems," in IEEE Transactions on Communications, vol. 67, no. 5, pp. 3768-3781, May 2019.

\bibitem{multier}
I. Bor-Yaliniz and H. Yanikomeroglu, "The New Frontier in RAN Heterogeneity: Multi-Tier Drone-Cells," in IEEE Communications Magazine, vol. 54, no. 11, pp. 48-55, November 2016.

\bibitem{hetopt}
P. Lohan and D. Mishra, "Utility-Aware Optimal Resource Allocation Protocol for UAV-Assisted Small Cells With Heterogeneous Coverage Demands," in IEEE Transactions on Wireless Communications, vol. 19, no. 2, pp. 1221-1236, Feb. 2020.

\bibitem{cachingslice}
S. Zhang, W. Quan, J. Li, W. Shi, P. Yang and X. Shen, "Air-Ground Integrated Vehicular Network Slicing With Content Pushing and Caching," in IEEE Journal on Selected Areas in Communications, vol. 36, no. 9, pp. 2114-2127, Sept. 2018, 

\bibitem{UAVassisted}
G. K. Xilouris, M. C. Batistatos, G. E. Athanasiadou, G. Tsoulos, H. B. Pervaiz and C. C. Zarakovitis, "UAV-Assisted 5G Network Architecture with Slicing and Virtualization," 2018 IEEE Globecom Workshops (GC Wkshps), Abu Dhabi, United Arab Emirates, 2018, pp. 1-7

\bibitem{tquek}
P. Yang, X. Xi, J. Chen, T. Q. S. Quek, X. Cao and D. O. Wu, "RAN Slicing in a UAV Network for eMBB Service Provision." arXiv preprint arXiv:1912.03600 (2019).

\bibitem{pfab}
M. M. Azari, G. Geraci, A. Garcia-Rodriguez and S. Pollin, "Cellular UAV-to-UAV Communications," 2019 IEEE 30th Annual International Symposium on Personal, Indoor and Mobile Radio Communications (PIMRC), Istanbul, Turkey, 2019, pp. 1-7


\end{thebibliography}
\end{document}